# Coherence times of $Ce^{3+}$ spin states in $CaWO_4$ crystal

M.R. Gafurov [1], I.N. Kurkin [1], E.I. Baibekov [1, *]

[1] Kazan Federal University, Kremlevskaya 18, Kazan 420008, Russia
*E-mail: edbaibek@gmail.com

We study the coherence times and perform manipulations on the lowest-energy states of trivalent cerium ion in calcium tungstate crystal. We find the phase memory time reaching 14.2 μs and the time of coherent manipulations reaching 0.3 μs in the low-temperature limit, the latter can potentially be elongated by using the rotation angle and off-resonance error correction schemes.

## 1. Introduction

Recently, impurity rare earth (RE) ions in inorganic crystals have attracted interest as spin qubits in quantum information processing [1-3]. Among all rare earth elements, only cerium lacks nuclear spin (even isotopes $^{136}$Ce, $^{138}$Ce, $^{140}$Ce and $^{142}$Ce with total natural abundance of 100%). Containing only one electron on the valence $4f$ shell, a trivalent cerium ion is a good representation of a single spin-1/2 particle. Coherent manipulation of $Ce^{3+}$ spin states in YAG has been studied recently [3,4]. Using dynamical decoupling, the phase memory time $T_2$ of cerium spin states reached 2 ms and were limited only by the spin-lattice relaxation time $T_1$ [4]. However, the characteristic times of coherent spin manipulation were found to be only < 2 μs. This so-called Rabi time $\tau_R$ characterizes the decay of spin nutations (or Rabi oscillations), which are excited by the continuously applied microwave (mw) pulse with the frequency in resonance with the spin transition [5]. The Rabi time limits the coherence exactly during the spin transition, and thus determines the fidelity of the one-qubit operation. This is in contrast to $T_2$ time, which govern the coherence of the spin qubit states between the operations. Together, these two times can characterize the one-qubit coherence in quantum information processing devices.

Over the last decades, RE-doped calcium tungstate crystals have attracted much interest for their applications in optical devices and quantum electronics [6]. Electron paramagnetic resonance (EPR) spectra of $CaWO_4$:$RE^{3+}$ crystals have been extensively studied, starting from the pioneering works of Mims and co-workers [7-10]. The titled compound is the one in which the electron spin echo envelope modulation was first observed in 1965 [11]. Calcium tungstate host crystal contains mostly ions without nuclear spin (with the exception of $^{183}$W and $^{43}$Ca with natural abundance of 14.3% and 0.135%, respectively). Thus, $T_2$ times of the electronic states of the impurity RE ions in $CaWO_4$ are sufficiently long with respect to the ones, e.g., in fluoride crystals, the former sometimes exceeding 0.1 ms [2].

While there is sufficient data on phase and spin-lattice relaxation of $Ce^{3+}$ in $CaWO_4$ crystal, the coherent manipulations of cerium electronic states have not been attempted yet. The aim of the present article is to demonstrate the possibility of coherent control over the two lowest states of $Ce^{3+}$ ions in $CaWO_4$ crystal, which model an ensemble of identical spin qubits hosted in almost nonmagnetic background.

## 2. EPR spectra, phase memory and spin-lattice relaxation times of $Ce^{3+}$ ion in $CaWO_4$ crystal

*Sample characterization and EPR spectra*

The sample of $CaWO_4$:$Ce^{3+}$ single crystal was grown by Czochralski method in the Magnetic Resonance Laboratory of Kazan Federal University by N. A. Karpov. Nominal concentration of cerium ions equaled 0.01 at. % ($C = 1.28 \cdot 10^{18}$ ions per c.c.). In order to reduce the effect of mw field inhomogeneity, a small plate with approximate dimensions of 2.4×1.4×2.6 mm was cut from the original sample. A quartz tube containing this smaller sample was placed close to the center of an X-band sapphire loop-gap resonator. The static magnetic field $\mathbf{B}_0$ was directed perpendicular to the crystal *c* axis. All measurements were performed using X-band Elexsys 580 EPR spectrometer at the mw frequency 9.69 GHz.

Continuous-wave EPR spectra of the crystal sample recorded at temperature $T$=20 K are shown in figure 1. In calcium tungstate, trivalent RE ions mostly substitute $Ca^{2+}$ at positions with tetragonal symmetry ($S_4$ point group). Most of the signal comes at $B_0 = 4842$ G (FWHM of 13 G), which corresponds to the transition with the *g*-factor $g = 1.430$. This is to be compared with the previously published data on *g*-factors of the ground Kramers doublet of the tetragonal center ($g_\perp = 1.423$, $g_\parallel = 2.915$) [12,13], and corresponds to the almost perpendicular orientation of the crystal *c* axis with respect to $\mathbf{B}_0$. Weaker and broader (FWHM of approximately 35 G) lines near 3370, 3577, 3644, 4010 and 4124 G most probably originate from the orthorhombic cites of $Ce^{3+}$ [14]. Comparing the total intensity of these signals with that of tetragonal centers, we estimate the relative concentration of these low-symmetry centers in our crystal sample to be ~ 10%.

*Phase relaxation*

In order to characterize the paramagnetic centers in our crystal sample, we performed measurements of $T_2$ and $T_1$ times of the tetragonal center. Phase memory time was recorded using Hahn spin echo pulse sequence $\pi/2 - \tau - \pi - \tau - \text{echo}$, where the interval $\tau$ between the pulses was incremented. In all spin echo measurements, the durations of $\pi/2$ and $\pi$ pulses where kept 16 ns and 32 ns, respectively. The echo intensity as a function of $\tau$ was fitted by a single exponential function $A\exp(-2\tau/T_2)$. Temperature dependence $T_2(T)$ of tetragonal centers is presented in tab. 1 and figure 2.

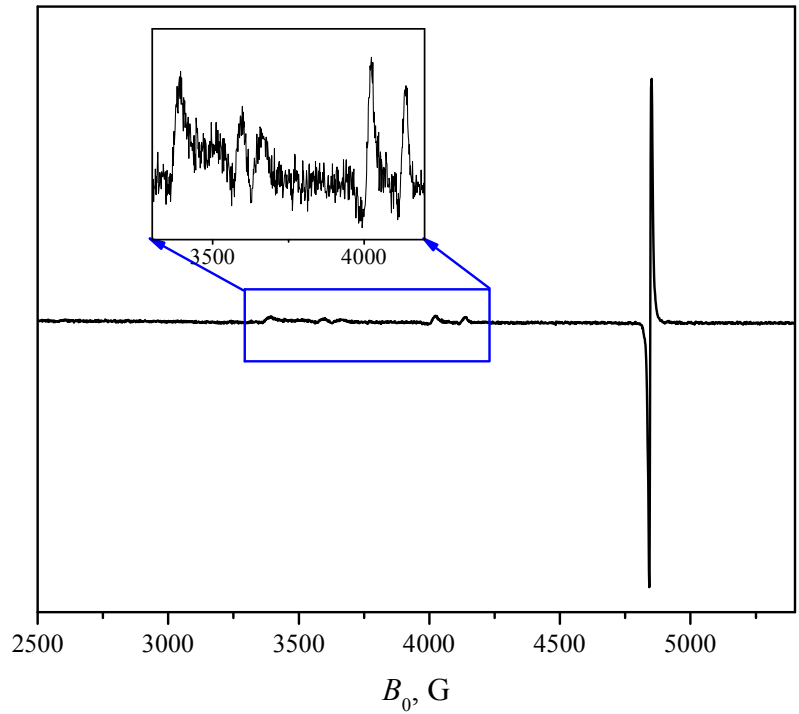

**Figure 1.** EPR spectrum of $Ce^{3+}$ ion in $CaWO_4$ crystal, $\mathbf{B}_0 \perp c$, $T = 20$ K. An inset shows the area of weaker signal corresponding to orthorhombic sites of $Ce^{3+}$.

Since the concentration of cerium ions was quite small, the main contribution to the transverse relaxation was magnetic dipole interaction between different $Ce^{3+}$ spins. These interactions contribute to the broadening of the EPR line with the half-width [10] $\Delta\omega_d = 2.53 g^2 \mu_B^2 C \hbar^{-1} = 5.4 \cdot 10^5$ s$^{-1}$ (or approximately 0.04 G), where $\mu_B$ is Bohr magneton, which is much less than the observed broadening. As a rule, EPR line of impurity ions is inhomogeneously broadened due to slightly different crystal field (nearby or distant charge compensators, lattice defects, strains, etc.). We take into account three different contributions into the transverse relaxation rate – instantaneous and spectral diffusion, and the spin-lattice relaxation:

$$T_2^{-1} = \Gamma_{ID} + \Gamma_{SD} + T_1^{-1}. \qquad (1)$$

The first one is instantaneous diffusion [15], and this is the only one independent of *T* in the studied temperature range. Here, flips of background cerium spins caused by the $\pi/2$ and $\pi$ pulses of the Hahn spin echo sequence induce fluctuations of local magnetic field acting on a given cerium ion. The corresponding contribution is $\Gamma_{ID} = \kappa \Delta\omega_d \langle \sin^2(\theta_\pi/2) \rangle$, where the expression in brackets $\langle ... \rangle$ is averaged over different spin packets within the same resonance line, $\theta_\pi$ is the angle by which a given spin packet is rotated by the $\pi$ pulse [15]. An additional factor $\kappa = 0.9$ is introduced in the above expression in order to exclude nonresonant transitions involving low-symmetry cerium centers. Only those packets detuned from the resonance frequency by less than Rabi frequency of the mw pulse, $\Omega_R = g_1 \mu_B B_1/\hbar$ (here $B_1$ is the amplitude of the mw field, $g_1$ is the g-factor along the direction of **B**$_1$), contribute effectively to the fluctuating field. Our calculations assuming Lorenzian line shape of the EPR line at 4842 G give $\langle \sin^2(\theta_\pi/2) \rangle = 0.7$ and $\Gamma_{ID}^{-1} = 2.9$ μs, which represents the low-temperature approximation of the phase memory time. This is to be compared with the experimental value of $T_2 \approx 14.3$ μs below 7 K. Since it is highly unlikely that there are other processes that reduce the instantaneous diffusion in our sample, we propose that the observed discrepancy originates from overestimated cerium impurity concentration. Indeed, since the substitution $Ca^{2+} \rightarrow Ce^{3+}$ is heterovalent, it requires charge compensation, which is not always present in sufficient quantities during the process of the crystal growth. Thus, actual concentration of RE impurity is usually smaller than the nominal one. In order to conform to the experimental low-temperature value of $T_2$, we estimate actual concentration of $Ce^{3+}$ ions to be 0.002 at. % ($C = 2.56 \cdot 10^{17}$ ions per c.c.), or 20% of the nominal concentration. This value is substantiated further by our calculation of temperature dependence $T_2(T)$ and will be used throughout the paper, including the next section (Rabi oscillations).

The second term in the Eq. (1) represents the contribution of spectral diffusion. Here, the local magnetic field fluctuations are induced by the spin flips of all background spins unrelated to the action of the mw pulses. In the case of small spin concentration and strong inhomogeneous broadening of the line, one can neglect spin flip-flops occurring between the spins of the same resonance frequency. In this case, the flip rate equals spin-lattice relaxation rate $T_1^{-1}$, and $\Gamma_{SD} = \sqrt{\Delta\omega_d T_1^{-1}}/2$ [15]. This process becomes dominant in the middle temperature range 9<$T$<15 K.

The last contribution in (1) is a simple spin-lattice relaxation, which dominates the phase relaxation above 15 K. Our calculations with corrected spin concentration (the third column of tab. 1) are in satisfactory agreement in the whole studied temperature range.

*Spin-lattice relaxation*

Spin-lattice relaxation time was obtained through an inversion recovery pulse sequence $\pi-\tau'$ followed by the spin echo detection sequence $\pi/2-\tau-\pi-\tau-\text{echo}$. There, $\tau$ was kept constant and $\tau'$ incremented with the minimal step of 4 ns. The results in the temperature range 5≤$T$≤18 K are depicted in figure 3.

**Table 1.** Temperature dependence of phase memory and spin-lattice relaxation time of $Ce^{3+}$ ion in $CaWO_4$ crystal. The values of $T_2$ in the third column are calculated according to Eq. (1) using a corrected cerium concentration, see text.

| $T$, K | $T_2$, µs experimental | $T_2$, µs calculated | $T_1$, µs |
|---|---|---|---|
| 5 | 14.2 | 14.3 | 24700 |
| 6 | 14.3 | 13.9 | 3100 |
| 7 | 14.8 | 12.9 | 550 |
| 8.1 | 9.9 | 11.0 | 129 |
| 9 | 8 | 8.1 | 37 |
| 10 | 5.3 | 5.4 | 14.1 |
| 11 | 3.4 | 3.5 | 6.6 |
| 11.5 | 2.9 | - | - |
| 12.5 | 1.81 | - | - |
| 14 | 0.74 | - | - |
| 15 | 0.47 | 0.48 | 0.56 |
| 16 | 0.295 | 0.282 | 0.314 |
| 18 | 0.117 | 0.144 | 0.155 |

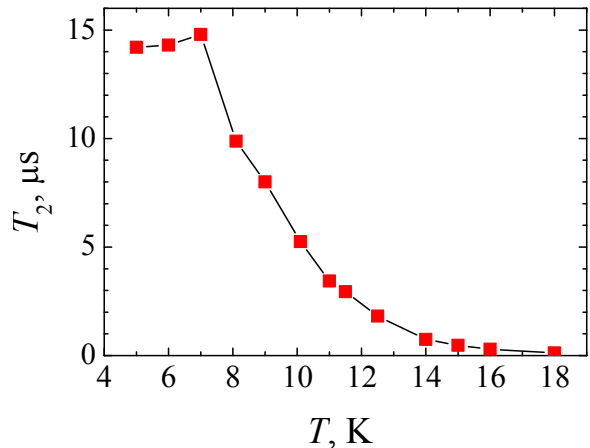

**Figure 2.** Temperature dependence of phase memory time of $Ce^{3+}$ ion in $CaWO_4$ crystal. Experimental error lies within the symbol size. Lines connecting experimental points are added to guide eye and are not experimentally verified.

The spin-lattice relaxation of $Ce^{3+}$ in $CaWO_4$ has been studied previously [16,17]. In the present section, we attempt to refine the literature data by extending the studied temperature range to 18 K. In ref. [16], the relaxation rate $T_1^{-1}(T)$ in the range 1.8 to 9 K was modelled by direct and Raman processes, the latter dominating above 4 K. Our data in the range 5≤T≤9 K are well fitted by the Raman-like dependence $B \cdot T^n$, where $n = 10.9$, in perfect agreement with the results presented previously [16]. However, in the range 10 to 18 K, the temperature dependence is better described by another power dependence $C \cdot T^{n'}$, with $n' = 7.8$. Note that we were unable to fit the dependence to the single-exponential Orbach-Aminov transition rate $\sim \exp(-\Delta/k_B T)$. It is unlikely that an Orbach process provides significant contribution below 20 K, since Raman spectroscopic data indicate that the next excited Kramers doublet lies above $\Delta > 100$ cm$^{-1}$ from the ground doublet [18]. At higher temperatures, the dependence starts to deviate from a simple power law. Above 18 K, the relaxation becomes too fast to be measurable by our experimental equipment. While it is well known that the Raman process contributes to the spin-lattice relaxation with $n$=9 [19], simple estimates show that it is naive to expect a good fit to $T^9$ for RE ions at temperatures above 4 K [16].

### 3. Rabi oscillations

Rabi oscillations of tetragonal cerium centers at 5 K were recorded by the following pulse sequence: mw pulse $t$ – delay – $\pi/2 - \tau - \pi - \tau - $ echo. There, the nutation pulse $t$ was followed by a delay of 100 µs, which was longer than $T_2$ but shorter than $T_1$, in order to get rid of the transverse magnetization component. This component was then measured by Hahn spin echo sequence. Then the whole sequence was repeated with an incremented $t$.

The experimental data recorded at various strengths of the mw field are shown by symbols in figure 4(a-d). Rabi frequencies $\Omega_R/2\pi$ were determined directly as the frequencies of the observed oscillations. The oscillations decay on a timescale ~ 1 μs, which is much less than the phase memory time of 14.2 μs observed at the same temperature. This indicates contribution of nonrelaxational processes, namely, spin dephasing due to (i) inhomogeneous broadening of the resonance line [20] and (ii) inhomogeneity of $B_1$ field inside the crystal sample [21]. A cerium spin detuned from the resonance frequency by $\varepsilon$ gives at time $t$ the following contribution to the crystal's magnetization component parallel to $\mathbf{B}_0$ field [20]:

$$s_z(\varepsilon, t) \sim \frac{\Omega_R^2 \cos(\Omega t) + \varepsilon^2}{\Omega^2}, \qquad (2)$$

where $\Omega = \sqrt{\Omega_R^2 + \varepsilon^2}$ is the nutation frequency of the spin. Since the mw field amplitude $B_1(\mathbf{r})$ is spatially distributed within the crystal sample, and so does $\Omega_R(\mathbf{r})$, one needs to integrate over the sample volume $V$. Time-dependent part of the crystal's magnetization becomes

$$M_z(\varepsilon, t) \sim \int d\varepsilon\, g(\varepsilon) \int_V dV\, \frac{\Omega_R^2(\mathbf{r}) \cos(\Omega(\mathbf{r}) t)}{\Omega^2(\mathbf{r})}. \qquad (3)$$

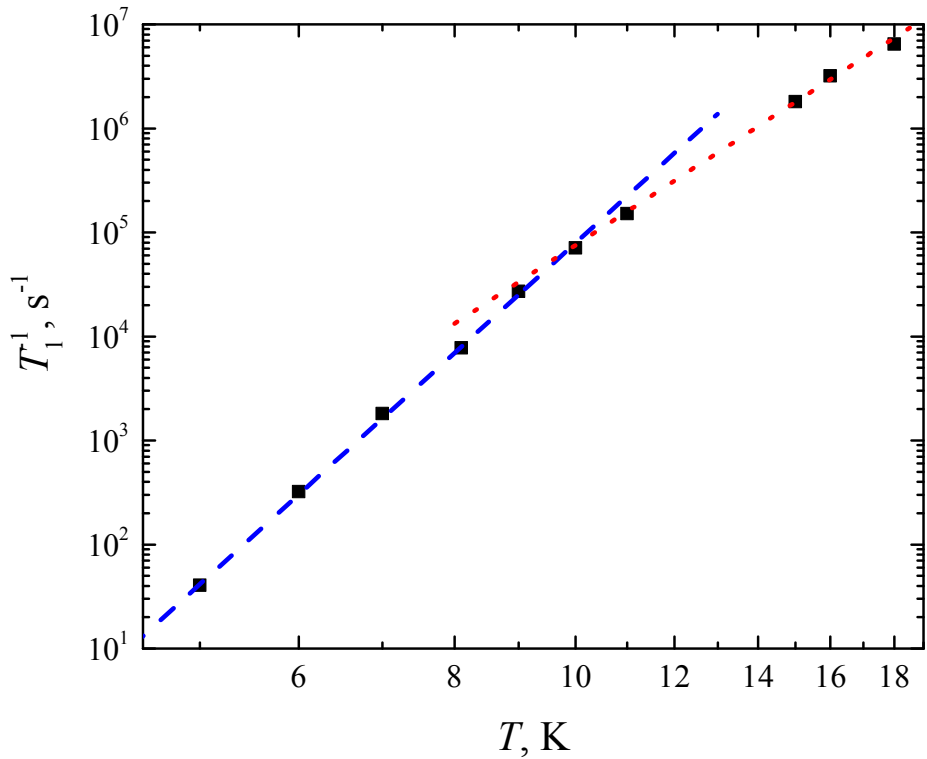

**Figure 3.** Temperature dependence of spin-lattice relaxation time of $Ce^{3+}$ ion in $CaWO_4$ crystal. Dashed and dotted lines represent power-law approximations $\sim T^{10.9}$ and $\sim T^{7.8}$, see text. Experimental error lies within the symbol size.

In the case of small mw field inhomogeneity, the averaging over $\mathbf{r}$ and integration over $\varepsilon$ within the EPR line governed by the distribution density $g(\varepsilon)$ can be performed separately. When the center of the sample is located at the antinode of the standing wave of the mw resonator, one can introduce a reasonable approximation to Eq. (3) [21]:

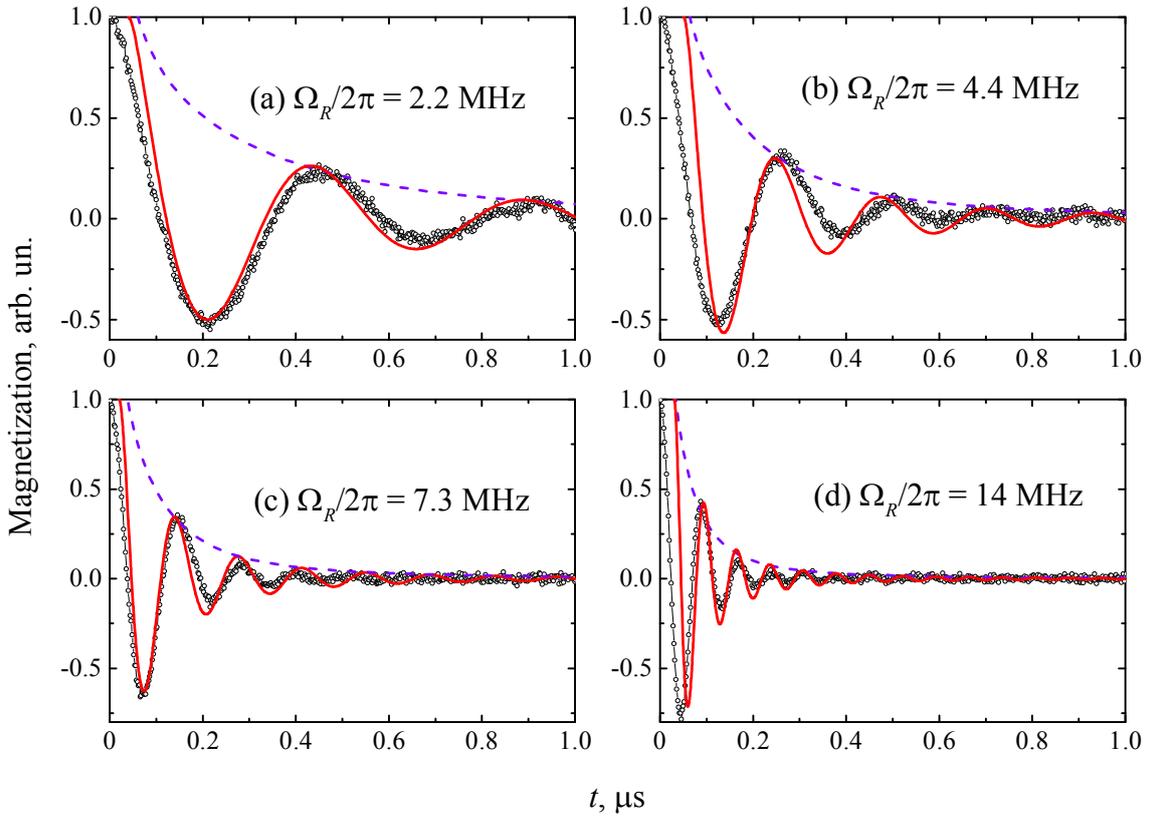

**Figure 4.** Rabi oscillations of Ce³⁺ ion in CaWO₄ crystal recorded at various strengths of the mw field corresponding to the nominal attenuation of the maximal applied mw power 18 (a), 12 (b), 6 (c) and 0 dB (a). $T$ = 5 K. Experimental data and the results of the calculations using Eqs. (4) and (6) are represented by symbols, red solid and blue dashed lines, respectively.

$$M_z(\varepsilon,t) \sim \left(1+(\beta_{B1}\Omega_R t)^2\right)^{-3/4} \int d\varepsilon\, g(\varepsilon) \int_V dV\, \frac{\Omega_R^2(0)\cos(\Omega(0)t)}{\Omega^2(0)}, \qquad (4)$$

where $\beta_{B1} \ll 1$ is the dimensionless parameter that determines the inhomogeneity of the mw field inside the sample: $1-\beta_{B1} \leq \Omega_R(\mathbf{r})/\Omega_R(0) \leq 1$, and $\Omega_R(0)$ is the Rabi frequency at the antinode (which we will further denote simply by $\Omega_R$). When the line half-width exceeds $\Omega_R$, one can simplify the Eq. (4) further as [21]

$$M_z(\varepsilon,t) \approx M_0(t)\cos(\Omega_R t + \varphi), \qquad (5)$$

where the amplitude of nutations decays as

$$M_0(t) \sim \left(1+(\pi\Omega_R t)^2\right)^{-1/4} \left(1+(\beta_{B1}\Omega_R t)^2\right)^{-3/4}. \qquad (6)$$

During the first period of spin nutations, the main contribution to the decay comes from the dephasing of various spin packets (the first factor in Eq. (6)). At higher $t$, mw field inhomogeneity becomes dominant. Ultimately, the amplitude decay is governed by a power law $M_0(t) \sim t^{-2}$. Note that, according to our results, the experimental decay curves are essentially nonexponential.

Magnetic dipole interactions between cerium spins result in an additional decay with the rate $\Gamma_d \leq \Delta\omega_d/2$ [22]. At the spin concentration of 0.002 at. %, this contribution is significant only at $t > 10$ μs and can be neglected on the studied timescale. The calculations by Eq. (4) using experimental lineshape $g(\varepsilon)$ give the best fit for $\beta_{B1} = 0.15$, see figure 4 for comparison with the experimental Rabi curves. Thus, we estimate that $B_1$ falls by 15% at the sample edge with respect to its maximum value at the antinode. The nutation amplitudes calculated by the Eq. (6) are drawn by the dashed lines.

The decay time $\tau_R$ of Rabi oscillations was estimated as moments by which the experimental amplitudes reduced by $e$ times. The decay rate $\tau_R^{-1}$ is plotted against $\Omega_R$ in figure 5. These data can be fitted by a simple linear dependence [21,23] $\tau_R^{-1} = 0.084\Omega_R + 2.7\cdot10^6$ s$^{-1}$.

In order to elongate the Rabi time, one needs to either reduce or compensate the spatial inhomogeneity of the mw field, as well as the broadening of the resonance line. The first task can be achieved by using smaller samples. However, as this would reduce the absolute number of cerium spins in the sample, the nutation signal would fall below the threshold of sensitivity of our spectrometer. Instead, one can introduce rotation angle and off-resonance error correction sequences [24], of which the most known are BB1 [25] and CORPSE [26], respectively. Ultimately, if one gets rid of the dephasing, the decay rates will approach the dipolar limit $\tau_R^{-1} \simeq \Delta\omega_d/2$, which, for our particular sample, equals $\tau_R = 20\,\mu s$.

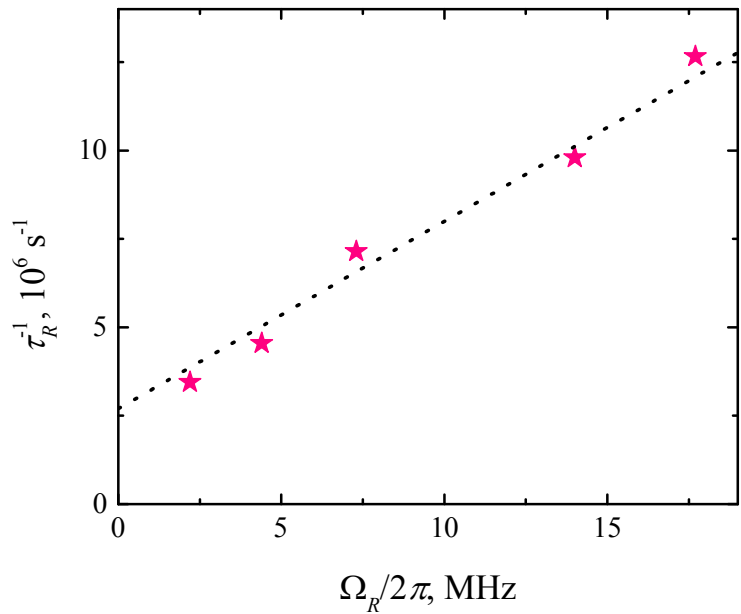

**Figure 5.** Decay rate of Rabi oscillations as a function of Rabi frequency. Experimental data and the linear fit are represented by asterisks and dashed line, respectively.

## 4. Conclusion

In the present article, we have demonstrated the coherent control over the lowest-energy Kramers states of $Ce^{3+}$ ion in $CaWO_4$ crystal. Using appropriate error correction sequences, one can potentially elongate the Rabi time to ~ 20 μs and perform up to 600 successive one-qubit NOT gates, each corresponding to a single 32-nanosecond $\pi$ pulse. We have also refined the existing data on phase and spin-lattice relaxation of the studied compound in the temperature range of 5 to 18 K. The results indicate that the selected material can be used as a model single-qubit system to test basic concepts of electron spin quantum computing.

## Acknowledgments

This work was financially supported by the Russian Science Foundation (Project no. 17-72-20053).

## References

1. Bertaina S., Gambarelli S., Tkachuk A., Kurkin I.N., Malkin B., Stepanov A., Barbara B. *Nature nanotechnology* **2**, 39 (2007)
2. Rakhmatullin R.M., Kurkin I.N., Mamin G.V., Orlinskii S.B., Gafurov M.R., Baibekov E.I., Malkin B.Z., Gambarelli S., Bertaina S., Barbara B. *Phys. Rev. B* **79**, 172408 (2009)
3. Azamat D.V., Belykh V.V., Yakovlev D.R., Fobbe F., Feng D.H., Evers E., Jastrabik L., Dejneka A., Bayer M. *Phys. Rev. B* **96**, 075160 (2017)
4. Siyushev P., Xia K., Reuter R., Jamali M., Zhao N., Yang N., Duan C., Kukharchyk N., Wieck A.D., Kolesov R., Wrachtrup J. *Nature Communications* **14**, 3895 (2014)
5. Rabi I.I. *Phys. Rev.* **51**, 652 (1937)
6. Kaminskii A.A., *Laser Crystals: Their Physics and Properties*, Springer-Verlag, Berlin (1981)
7. Mims W.B. *Phys. Rev.* **133**, A835 (1964)
8. Mims W.B. *Phys. Rev.* **140**, A531 (1965)
9. Mims W.B., Gillen R. *The Journal of Chem. Phys.* **47**, 3518 (1967)
10. Mims W.B. Phys. Rev. **168**, 370 (1968)
11. Rowan L.G, Hahn E.L., Mims W.B. *Phys. Rev.* **137**, A61 (1965)
12. Kurkin I.N., Shekun L. Ya., *Sov. Phys. Solid State* **7**, 2308 (1965)
13. Yang M., Wen-Chen Z. *Physica B: Cond. Matt.* **416**, 8 (2013)
14. Volterra V., Bronstein J., Rockni M. *Appl. Phys. Lett.* **8**, 212 (1966)
15. Salikhov K.M., Dzuba S.A., Raitsimring A.M. *J. Magn. Res.* **42**, 255 (1981)
16. Kiel A., Mims W.B. *Phys. Rev.* **161**, 386 (1967)
17. Antipin A.A., Katyshev A.N., Kurkin I.N., Shekun L.Y. *Sov. Phys. Solid State* **10**, 1136 (1968)
18. Chau J.Y.H. *The Journal of Chem. Phys.* **44**, 1708 (1966).
19. Abragam A., Bleaney B. *Electron Paramagnetic Resonance of Transition Ions*, Clarendon Press, Oxford (1970)

20. Torrey H.C. *Phys. Rev.* **76**, 1059 (1949)
21. Baibekov E., Kurkin I., Gafurov M., Endeward B., Rakhmatullin R., Mamin G. *J. Magn. Res.* **209**, 61 (2011)
22. Baibekov E.I. *JETP Lett*. **93**, 292 (2011)
23. Shakhmuratov R.N., Gelardi F.M., Cannas M. *Phys. Rev. Lett.* **79**, 2963 (1997)
24. Ishmuratov I.K., Baibekov E.I. *J. Low Temp. Phys.* **185**, 583 (2016)
25. Wimperis S. *J. Magn. Res.* **109**, 221 (1994)
26. Alway W.G., Jones J.A. *J. Magn. Res.* **189**, 114 (2007)